\documentclass[aps,prl,preprint,superscriptaddress,showpacs]{revtex4-1}
\usepackage{graphics}
\usepackage[english]{babel}
\usepackage{epstopdf}
\usepackage{amsmath}
\usepackage{amsfonts}
\usepackage{amssymb}
\usepackage[latin1]{inputenc} %caratteri speciali
\usepackage[OT2,T1]{fontenc}

\begin{document}

%\preprint{}

\title{Thick Permalloy films for the imaging of spin texture dynamics in perpendicularly magnetized systems}


%%%% AUTHOR LIST

\author{S. Finizio}
%\thanks{}
\email[Corresponding Author: ]{simone.finizio@psi.ch}
\affiliation{Paul Scherrer Institut, 5232 Villigen PSI, Switzerland}

\author{S. Wintz}
\affiliation{Paul Scherrer Institut, 5232 Villigen PSI, Switzerland}
\affiliation{Institute of Ion Beam Physics and Materials Research, Helmholtz-Zentrum Dresden-Rossendorf, 01328 Dresden, Germany}

\author{D. Bracher}
\affiliation{Paul Scherrer Institut, 5232 Villigen PSI, Switzerland}

%\author{M. Mruczkiewicz}
%\affiliation{Institute of Electrical Engineering, Slovak Academy of Sciences, 841 04 Bratislava, Slovak Republic}

\author{E. Kirk}
\affiliation{Paul Scherrer Institut, 5232 Villigen PSI, Switzerland}
\affiliation{Laboratory for Mesoscopic Systems, Department of Materials, ETH Z\"{u}rich, 8093 Z\"{u}rich, Switzerland}

\author{A. S. Semisalova}
\affiliation{Institute of Ion Beam Physics and Materials Research, Helmholtz-Zentrum Dresden-Rossendorf, 01328 Dresden, Germany}

\author{J. F\"orster}
\affiliation{Max-Planck-Institut f\"ur Intelligente Systeme, 70569 Stuttgart, Germany}

\author{K. Zeissler}
\affiliation{School of Physics and Astronomy, University of Leeds, LS2 9JT Leeds, United Kingdom}

\author{T. We{\ss}els}
\affiliation{Ernst Ruska-Centre for Microscopy and Spectroscopy with Electrons and Peter Gr\"unberg Institute, Forschungszentrum J\"ulich, 52425 J\"ulich, Germany}

\author{M. Weigand}
\affiliation{Max-Planck-Institut f\"ur Intelligente Systeme, 70569 Stuttgart, Germany}

\author{K. Lenz}
\affiliation{Institute of Ion Beam Physics and Materials Research, Helmholtz-Zentrum Dresden-Rossendorf, 01328 Dresden, Germany}

\author{A. Kleibert}
\affiliation{Paul Scherrer Institut, 5232 Villigen PSI, Switzerland}

\author{J. Raabe}
\affiliation{Paul Scherrer Institut, 5232 Villigen PSI, Switzerland}

%%%% END OF AUTHOR LIST
\date{\today}

%\doublespacing

\begin{abstract}
We demonstrated that thick Permalloy films exhibiting a weak growth-induced perpendicular magnetic anisotropy can be employed as an ideal test system for the investigation of gyration dynamics in topologically trivial and non-trivial magnetic states ranging from an isolated magnetic skyrmion to more complex $n\pi$ spin configurations.
\end{abstract}

% insert suggested PACS numbers in braces on next line`
%\pacs{68.37.Yz, 75.80.+q, 75.78.Cd}
% 68.37.Yz - X-ray microscopy
% 75.80.+q - Magnetostriction
% 75.78.Cd - Micromagnetic simulations

%\keywords{}

\maketitle

\section{Introduction}

%The study of materials exhibiting perpendicular magnetic anisotropy (PMA) has recently attracted a significant attention, mostly thanks to their many technologically relevant applications. Examples include current driven domain wall motion \cite{art:parkin_CIDWM}, ultrafast magnetic switching \cite{art:baumgartner_switching}, and magnetic skyrmions . 

The investigation of topological spin textures (such as e.g. the magnetic skyrmion \cite{art:nagaosa_skyrmion_topology, art:kai_skyrmion_hall_angle, art:hofmann_skyrmion_hall_angle, art:fert_topo_protection}) in material systems exhibiting perpendicular magnetic anisotropy (PMA) has recently been object of increased attention. This is due to the properties arising from their non-trivial topology due to their topological charge, where the topological charge is defined according to the following equation \cite{art:braun_topology}:
\begin{equation}
Q = \frac{1}{4\pi} \int{\mathbf{m} \cdot \left( \frac{\partial \mathbf{m}}{\partial x} \times \frac{\partial \mathbf{m}}{\partial y} \right) \mathrm{d}x \mathrm{d}y,}
\end{equation}
being $\mathbf{m}$ the normalized magnetization vector.

Some of the properties influenced by the topological charge include e.g. the topological Hall effect \cite{art:nagaosa_skyrmion_topology}, the skyrmion Hall effect \cite{art:nagaosa_skyrmion_topology, art:kai_skyrmion_hall_angle, art:hofmann_skyrmion_hall_angle}, and the topological protection of these entities, which provides an energy barrier against annihilation at defects and pinning sites \cite{art:fert_topo_protection, art:buettner_skyrmion_energy_barrier}. It is also predicted that the topological charge has a considerable influence over the magneto-dynamical processes (e.g. gyration dynamics) of these magnetic configurations \cite{art:moutafis_gyration, art:komineas_skyrmionium, art:makhfudz_gyration}, prompting for an experimental verification of such predictions. A verification of this kind usually relies on pump-probe magnetic imaging, combining a high spatial and temporal resolution. However, a fundamental requirement for these experiments is that the dynamical processes need to be reproducible over a number of excitation cycles on the order of 10$^6$-10$^{10}$. This comes with the requirement that the Gilbert damping of the magnetic material should be low, to allow one to excite the gyration modes with moderate excitation signals, and that the material should exhibit a low density of pinning sites.

The PMA materials typically employed for the investigation of magnetodynamical processes in topological magnetic configurations such as the magnetic skyrmion usually consist of NM1/FM/NM2 (FM: ferromagnet, NM: non-magnetic material) multilayer superlattice stacks optimized for a high PMA. Examples of these multilayer superlattices include Pt/Co/Pt \cite{art:mizukami_CoPt_damping}, Pt/Co/Ir \cite{art:zeissler_pinning}, Pt/CoFeB/MgO \cite{art:kai_skyrmion_hall_angle}, and W/CoFeB/MgO \cite{art:jaiswahl_WCoFeBMgO}. However, these multilayer superlattice stacks are usually afflicted by both a relatively high Gilbert damping (e.g. Pt/Co multilayer superlattices optimized for a high PMA show Gilbert dampings on the order of 0.2 \cite{art:barman_CoPt_damping, art:mizukami_CoPt_damping}), leading to short-lived dynamics \cite{art:buettner_gyration}, and a high density of pinning sites, which considerably influences the behavior of the magnetic configuration both statically \cite{art:zeissler_pinning, art:gross_pinning} and dynamically \cite{art:buettner_gyration}.

In this work, we propose an alternative approach to the use of multilayer PMA superlattice stacks for the time-resolved investigation of the dynamical processes in perpendicularly magnetized systems. This solution relies on the use of a much simpler material: Permalloy (Py - Ni$_x$Fe$_{1-x}$ alloy). Thin Py films have been, thanks to a combination of a low Gilbert damping and a relatively low density of pinning sites, one of the favorite systems for the investigation of magnetodynamical processes in in-plane magnetized systems. However, if the Py films are grown at thicknesses above a critical value, the presence of a weak, growth-induced, PMA leads to the stabilization of a worm-like perpendicularly magnetized stripe domain state \cite{art:iwata_PMA_Py, art:saito_PMA_Py, art:Lo_PMA_Py, art:eames_thickPy, art:youssef_PMA_Py, art:wei_PMA_Py}. This is observed within a relatively wide range of stoichiometries for the Ni$_x$Fe$_{1-x}$ alloy, suggesting that the origin for this effect is not only due to magnetostrictive effects \cite{art:saito_PMA_Py}. Furthermore, if microstructured discs are fabricated out of these thick Py films, magnetic states ranging from isolated magnetic skyrmions to $n\pi$ magnetic configurations (such as e.g. the 2$\pi$ state \cite{art:komineas_skyrmionium}) can be reliably stabilized by tailoring the diameter of the discs \cite{art:eames_thickPy}.

The presence of such magnetic states in thick Py films has been known since more than 50 years \cite{art:iwata_PMA_Py, art:saito_PMA_Py, art:Lo_PMA_Py, art:eames_thickPy, art:youssef_PMA_Py, art:wei_PMA_Py}, but, surprisingly, no experimental attempt to investigate the dynamical processes (such as e.g. the gyration dynamics of the magnetic skyrmions stabilized at the center of the disc structures) has been carried out. We then demonstrate in this work that the advantages that made thin Py films one of the favorite systems for the study of magnetodynamical processes are maintained also for the thick Py films presented here, allowing us to report on a first proof-of-principle measurement of the gyrotropic motion of a magnetic bubble domain in a thick Py microstructured disc, to demonstrate the feasibility of using thick Py films as a simple and reliable testbed for the investigation of magnetodynamical processes in topological spin textures.

\section{Experimental}

Microstructured thick Py (with Ni$_{81}$Fe$_{19}$ stoichiometry) disc elements (diameters ranging from 500 nm to 3 $\mu$m) were lithographically patterned on top of 200 nm thick x-ray transparent Si$_3$N$_4$ membranes and of p-doped Si(001) substrates. A bilayer of methyl-methacrylate and of poly-methyl-methacrylate was spincoated on top of the substrates prior to the lithographical exposure, which was carried out using a Vistec EBPG 5000Plus 100 kV electron beam writer. The structures were exposed with an exposure dose of 1500 $\mu$C cm$^{-2}$ for the Si$_3$N$_4$ membranes, and of 900 $\mu$C cm$^{-2}$ for the Si substrates, to account for the different scattering of the electrons from the different substrates. Following the lithographical exposure, the resist was developed by immersion in a solution of methyl-isobutyl-ketone and isopropanol (1:3 by volume) for 60 s, followed by immersion in pure isopropanol for 60 s.

The Py films were deposited by thermal evaporation from a commercial Ni$_{81}$Fe$_{19}$ pellet, with a growth rate of about 0.5 nm s$^{-1}$ (measured with a quartz crystal balance), using a Balzers BAE 250 evaporator with a base pressure in the 10$^{-7}$ mbar range. Prior to the deposition of the Py, a Cr adhesion layer of 10 nm was thermally evaporated on top of the substrates. After deposition, the Py films were capped with 5 nm of Cr to prevent oxidation. The thickness of the Py films was verified by atomic force microscopy on a reference sample. A thickness of 180 nm for the Py films was selected as a compromise between the necessity to obtain a weak PMA, which requires thick films \cite{art:saito_PMA_Py}, and the x-ray photon transmission across the Py, necessary for the scanning transmission x-ray microscopy (STXM) imaging, which requires thin films.

After the deposition of the Py films, the parts of the film grown on top of the unexposed resist areas were lifted off by immersion in pure acetone. The quality of the lifted-off structures was verified by optical and scanning electron microscopy.

The magnitude of the Gilbert damping and of the PMA of the continuous Py films was determined by broadband ferromagnetic resonance (FMR) measurements on equivalent reference samples.

The magnetic configuration of the microstructured Py elements was characterized by x-ray photoemission electron microscopy (PEEM) at the Surface Interface Microscopy (SIM) beamline \cite{art:SIM}, and by STXM at the PolLux (X07DA) endstation \cite{art:pollux}, both at the Swiss Light Source. Magnetic contrast in the resulting images was achieved through the x-ray magnetic circular dichroism (XMCD) effect \cite{art:schuetz_xmcd}. The circularly-polarized x-rays were tuned to the L$_3$ absorption edge of Fe.

\begin{figure*}
	\includegraphics{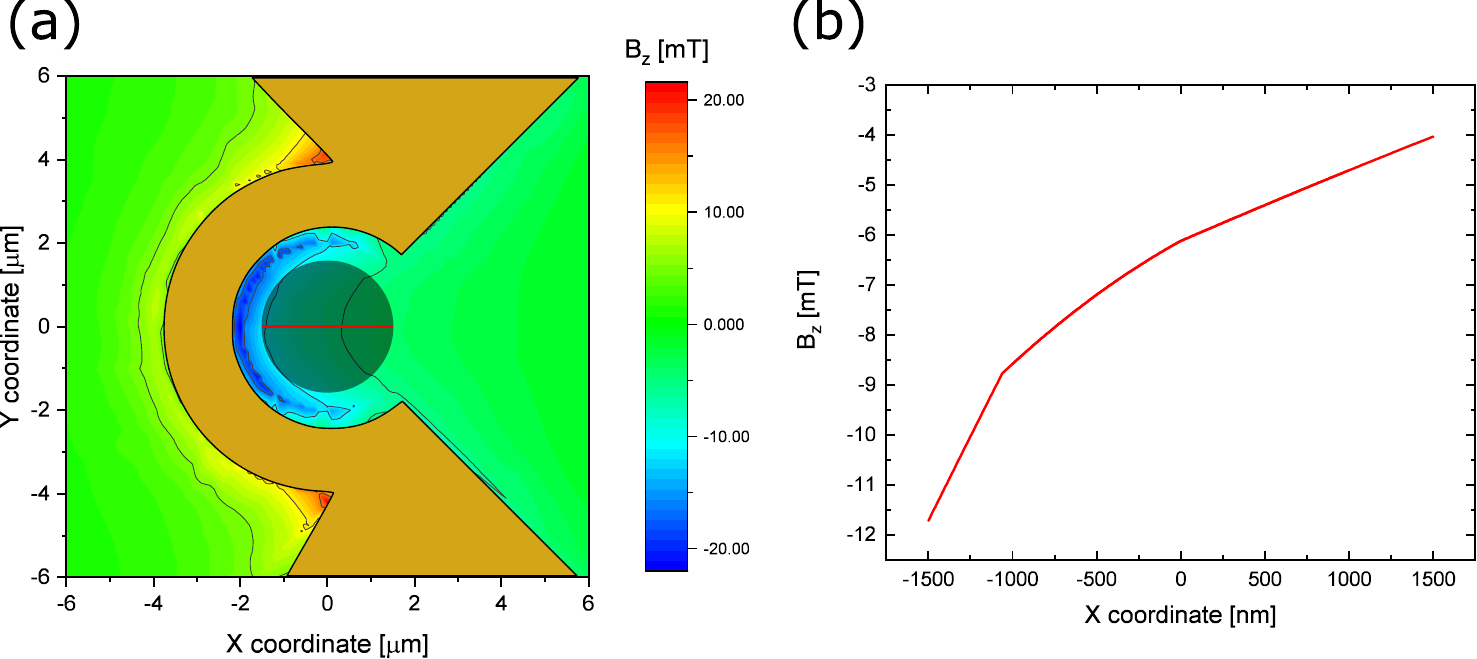}
	\caption{Finite element simulation of the out-of-plane magnetic field generated by the $\Omega$-shaped microcoil described in the main manuscript. (a) Finite element simulation of the z component of the magnetic field generated by the $\Omega$-shaped microcoil (with a 50 mA current injected across the coil). The circle marks the position of the 3 $\mu$m disc reported in the manuscript. (b) Amplitude of the z component of the magnetic field generated by the coil along the red line shown in (a). A clear gradient of the magnetic field across the disc can be observed.}
	\label{fig:SUP_omega_coil}
\end{figure*}

Thanks to the 16\textdegree incidence angle of the x-rays with respect to the surface of the sample, the XMCD-PEEM imaging experiments allowed for the investigation of the in-plane component of the magnetic domains in both the continuous films and in the microstructured elements fabricated on top of the doped Si substrates. In particular, the in-plane and out-of-plane spin configuration of the thick Py films were determined by acquiring XMCD-PEEM images of the sample under an azimuthal rotation of 0\textdegree and 180\textdegree, and subtracting/adding the two images. Further details on the technique are described in Ref. \cite{art:SIM}. The XMCD-PEEM images were acquired under no applied static external magnetic fields (referred to as remnant state). The spatial resolution for the XMCD-PEEM images presented here is on the order of 50 to 75 nm. Note that, due to the surface sensitivity of PEEM imaging \cite{art:schoenhense_peem_probing_depth}, only the magnetization configuration at the top surface of the Py films could be investigated with this technique.

To complement the results obtained from XMCD-PEEM imaging, the Py microstructured discs fabricated on top of x-ray transparent Si$_3$N$_4$ membranes were investigated by XMCD-STXM imaging. A Fresnel zone plate with an outermost zone width of 25 nm was employed to focus the circularly polarized x-rays. The entrance and exit slit to the monochromator were set in order to achieve a beam spot on the order of 25 nm. Due to the normal incidence of the x-ray beam with respect to the sample surface, only the out-of-plane component of the magnetization of the microstructured Py elements could be resolved with XMCD-STXM imaging. 

The response of the Py discs to static magnetic fields was determined with quasi-static XMCD-STXM, and time-resolved STXM imaging was employed to image the magnetization dynamics excited by an oscillating out-of-plane magnetic field gradient in the Py microstructured elements.  The time-resolved STXM imaging experiments were performed using circularly polarized photons of only one helicity (circular negative) through the pump-probe technique, as described in detail in Ref. \cite{art:puzic_TRSTXM}. The pump signal consisted of an oscillating out-of-plane magnetic field, generated by injecting an oscillating current across a Cu $\Omega$-shaped microcoil, generated with a Tektronix AWG7122C arbitrary waveform generator. The microcoil was lithographically defined to be 2 $\mu$m wide, and 200 nm thick, and was fabricated around a  $\mu$m diameter Py disc. To determine the intensity of the out-of-plane magnetic field gradient generated by the $\Omega$-shaped microcoil, finite element simulations were carried out with the commercial software suite ANSYS. The magnetic field was simulated with a current injected across the $\Omega$-shaped coil of 50 mA, which corresponds to the maximum of the applied current during the experiments. The value of the current was determined by measuring the current transmitted across the coil with a 50 $\Omega$-terminated real-time oscilloscope (Agilent DSO-S 404A). The results of the simulations are shown in Fig. \ref{fig:SUP_omega_coil}. A clear gradient in the out-of-plane component of the magnetic field can be observed in Fig. \ref{fig:SUP_omega_coil}(b).

The probing signal is given by the x-ray flashes generated (at a frequency of 500 MHz) from the synchrotron light source. The waveform generator is synchronized to the master clock of the synchrotron through a dedicated field programmable gate array setup, which also handles the timing for the acquisition of the time-resolved data. For the data presented in this manuscript, the time resolution was of about 200 ps.

\section{Static properties}

The thick Py films reported here exhibit a weak, growth-induced, PMA. The uniaxial anisotropy constant was determined from the value of the effective magnetization $\mu_0$M$_\mathrm{eff}$ according to the following relation:

\begin{equation}
\mu_0 \mathrm{M}_\mathrm{eff} \, = \, \mu_0 \mathrm{M}_\mathrm{s} - 2 \frac{\mathrm{K}_\mathrm{u}}{\mathrm{M}_\mathrm{s}},
\end{equation}

where M$_\mathrm{s}$ = 765.8 kA m$^{-1}$ is the saturation magnetization obtained from superconducting quantum interference device magnetometry. The value of $\mu_0 \mathrm{M}_\mathrm{eff}$ was determined from FMR measurements by fitting both the frequency and the polar angular dependencies of the resonance field. The PMA constant was measured to be of 29.6 kJ m$^{-3}$.

\begin{figure}
 \includegraphics{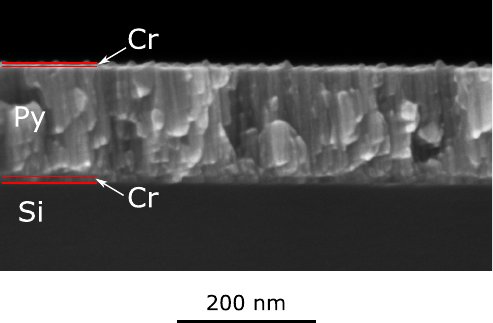}
 \caption{Cross-section scanning electron micrograph of a 180 nm thick Py film, with a 10 nm Cr adhesion layer and a 5 nm Cr capping layer, showing the columnar growth of the Py resulting in the weak PMA observed for these films.}
 \label{fig:SEM}
\end{figure}

This weak PMA is attributed to shape anisotropy effects caused by the columnar growth of the Py films \cite{art:Lo_PMA_Py, art:iwata_PMA_Py, art:saito_PMA_Py}. The columnar growth of the Py films presented here was qualitatively verified by scanning electron microscopy imaging of the cross-section of the as-grown Py films, an example of which is shown in Fig. \ref{fig:SEM}.

\begin{figure}
 \includegraphics{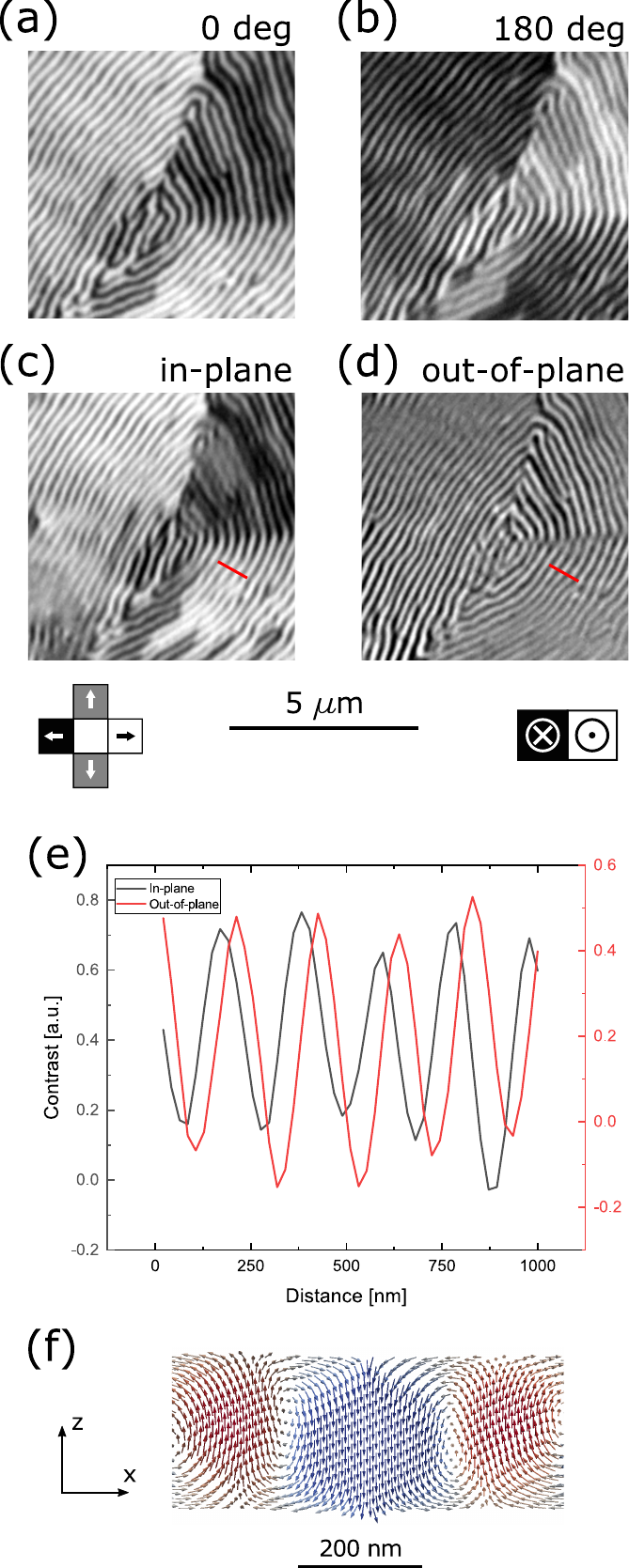}
 \caption{XMCD-PEEM images of a continuous, 180 nm thick, Py film (field of view 7.5 $\mu$m), showing (a-b) the original XMCD-PEEM images employed to determine (c) the in-plane (by subtracting the two images) and (d) the out-of-plane (by summing the two images) magnetic configuration of the Py. A stripe domain state, with a domain periodicity of about 250 nm is stabilized. The grayscale arrows in (c) and (d) sketch the direction of the magnetization deduced from the observed XMCD contrast. The original XMCD-PEEM images employed for calculating the in-plane and out-of-plane components of the magnetization are shown in the supplementary information. A linescan across the in-plane and out-of-plane images (marked in red in (c) and (d)) is shown in (e). The characteristic signature of N\'eel domain walls can be observed \cite{art:boulle_skyrmions}. (f) shows an overview of the spin configuration along the thickness of the Py film obtained from micromagnetic simulations.}
 \label{fig:PEEM}
\end{figure}

As shown in Fig. \ref{fig:PEEM}, the continuous, 180 nm thick, Py films stabilize a stripe domain pattern with a domain periodicity of about 250 nm. By measuring the in-plane and out-of-plane contrast across a number of different domains (Fig. \ref{fig:PEEM}(e)), it is possible to observe the signature of a N\'eel domain wall at the top surface of the film, similarly to the observations (on a different material) reported in Ref. \cite{art:boulle_skyrmions}. This observation is in agreement with the expected configuration of the magnetic domain wall for thick materials with a weak PMA, schematically depicted in Fig. \ref{fig:PEEM}(f), where it can be observed that the magnetic domain wall resembles a Bloch domain wall at the center of the film, and a N\'eel domain wall (of opposite chiralities) at the top and bottom surfaces of the Py film, also referred to as N\'eel closure caps \cite{art:moutafis_weakPMA, art:komineas_weakPMA, art:marioni_thickNi, art:eames_thickPy, art:wei_stripe_domains, art:durr_neel_caps}. It is worth to note here that the expected domain wall configuration for the thick Py films presented here resembles the one simulated for thick multilayer superlattice stacks exhibiting PMA and asymmetric exchange interaction, providing a further similarity between the spin configurations observed in the thick Py films and those observed in multilayered PMA superlattice stacks \cite{art:legrand_domainWall}.

%As shown in Fig. \ref{fig:PEEM}, and as reported in Ref. \cite{art:saito_PMA_Py}, the out-of-plane magnetic domain structure of the Py films investigated here is influenced by the in-plane component of the magnetic domain structure. In particular, it was reported that the worm domains align themselves parallel to the direction of the in-plane component of the domain structure \cite{art:saito_PMA_Py}. This observation is reproduced in the Py films investigated here, as shown in Fig. \ref{fig:Squares}(a). The perpendicularly-magnetized worm domains align themselves parallel to the edges of the microstructured squares, i.e. aligning themselves parallel to the in-plane component of the magnetization, which generates a flux-closure Landau pattern. In Fig. \ref{fig:Squares}(b), it is possible to observe that, if the microstructured Py elements exhibit a circular geometry, a magnetic state composed of a central bubble surrounded by concentric ring domains of opposite magnetizations will be stabilized.

%\begin{figure}
 %\includegraphics{Images/PEEM_squares.pdf}
% \caption{PEEM squares}
% \label{fig:Squares}
%\end{figure}

\begin{figure}
 \includegraphics{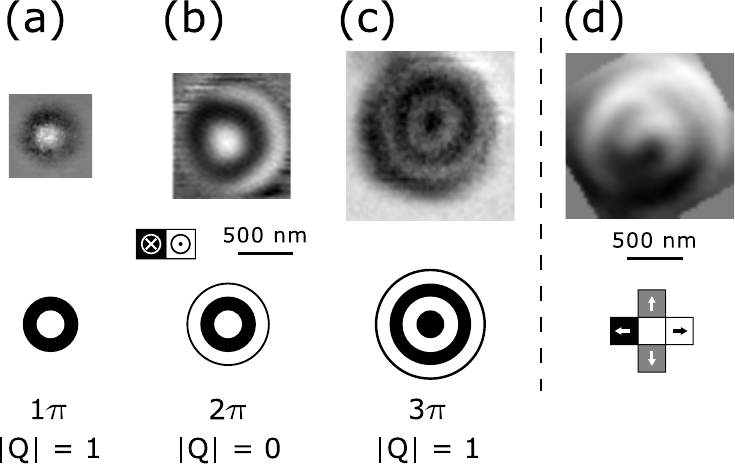}
 \caption{XMCD-STXM images of (a) a 1$\pi$ state (skyrmion) in a 500 nm diameter Py disc, (b) a 2$\pi$ state in a 750 nm diameter Py disc, and (c) a 3$\pi$ state in a 1 $\mu$m diameter Py disc. Below each disc, a schematic overview of the out-of-plane magnetic configuration is shown. Image (d) shows a corresponding XMCD-PEEM image of the in-plane component of a 1 $\mu$m diameter Py disc, where the N\'eel configuration of the domain walls can be observed. The grayscale arrows indicate the direction of the magnetic contrast.}
 \label{fig:STXM_vs_size}
\end{figure}

As shown in Fig. \ref{fig:STXM_vs_size}, if microstructured elements with a circular geometry are fabricated, thanks to the contribution of the shape anisotropy, a magnetic state composed of a central bubble surrounded by concentric ring domains of opposite magnetization will be stabilized \cite{art:eames_thickPy}. 

\begin{figure*}
	\includegraphics{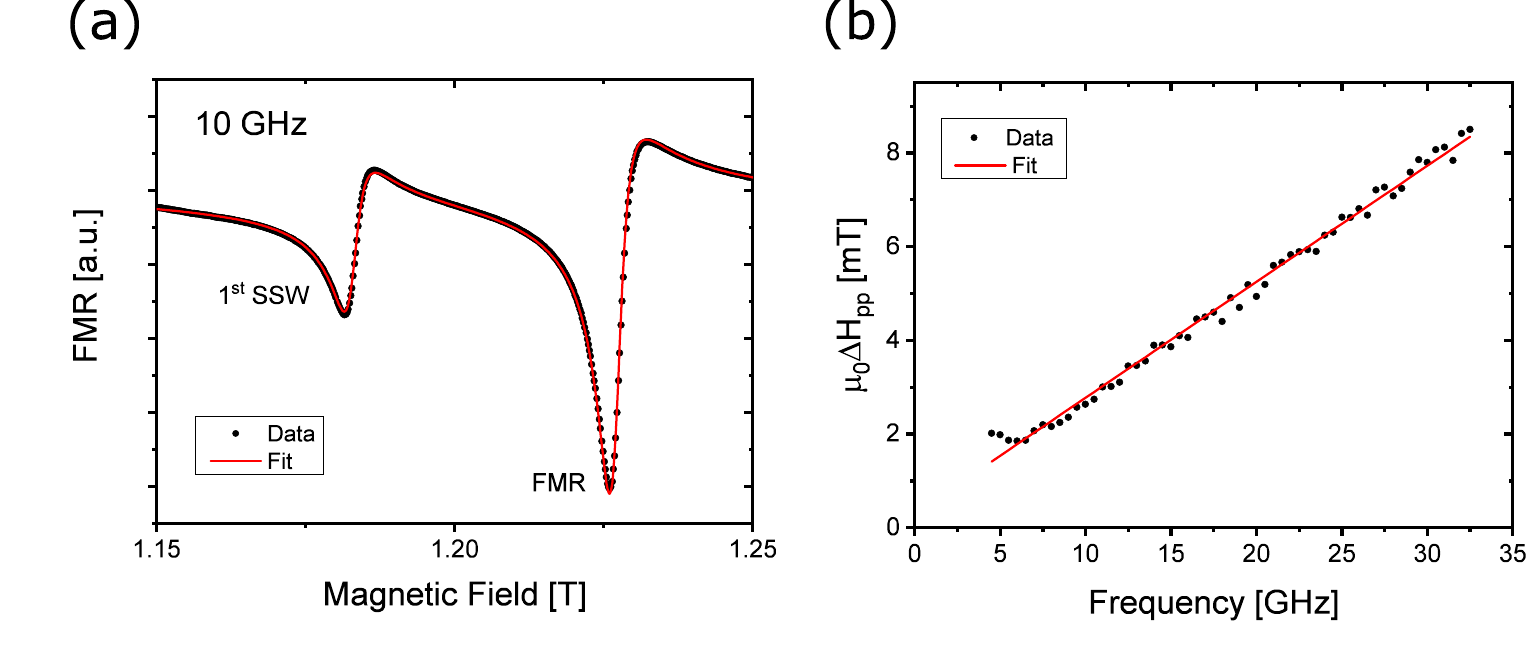}
	\caption{(a) Microwave absorption spectrum measured at a frequency of 10 GHz with a magnetic field applied along the out-of-plane direction. The fundamental FMR mode and the first standing spin wave mode are shown. The fit was performed using a complex Lorentzian. (b) Frequency dependence of the peak-to-peak linewidth of the fundamental FMR mode. The slope of the linear fit was used to extract Gilbert damping constant $\alpha$.}
	\label{fig:SUP_FMR}
\end{figure*}

Different magnetic configurations can be attained by tailoring the diameter of the microstructured disc elements. In particular, by selecting an integer multiple $N$ of 250 nm as the diameter of the microstructured discs, out-of-plane magnetic configurations ranging from an isolated skyrmion ($n=2$, see Fig. \ref{fig:STXM_vs_size}(a)), corresponding to a topological charge of $|Q| = 1$ to a central magnetic bubble surrounded by $n = (N-1)$ ring domains of alternating magnetization ($n\pi$ state - see Fig. \ref{fig:STXM_vs_size}(b)-(c) for the $N=3$ and $N=4$ examples, corresponding, respectively, to topological charges of $|Q| = 0$ and $|Q| = 1$) can be stabilized. Fig. \ref{fig:STXM_vs_size}(d) shows the in-plane component of a Py disc stabilizing a 3$\pi$ state ($N=4$), where it is possible to observe that the out-of-plane spin texture is coexisting with an in-plane vortex state. This observation is in agreement with previous works \cite{art:saito_PMA_Py}, where it was observed that the stripe domains align themselves parallel to the direction of this coexisting in-plane magnetic spin texture.

It is worth to mention here that the magnetic states shown in Fig. \ref{fig:STXM_vs_size} were acquired at the remnant state, i.e. in absence of any externally applied magnetic fields. This is in contrast with the majority of the PMA superlattice stacks employed for the stabilization of comparable spin configurations, where an out-of-plane magnetic field is necessary \cite{art:zeissler_pinning, art:kai_skyrmion_hall_angle, art:woo_skyrmions, art:buettner_gyration}.

Due to shape anisotropy, the selection of the diameter of the Py discs allows for the reliable and reproducible stabilization of magnetic states with different topological charges, ranging from the isolated skyrmion to the more complex $n\pi$ states. However, as already mentioned above, to study their magnetodynamical properties, a low Gilbert damping, and a low density of pinning sites are also desirable requirements.

The damping of the thick Py films presented here was measured by broadband FMR. The FMR spectra measured under an applied out-of-plane field at a frequency of 10 GHz are shown in Fig. \ref{fig:SUP_FMR}(a). A fundamental (uniform) mode and the first standing spin wave mode can be resolved from the FMR spectra. To allow for the determination of both the resonance field and its linewidth, the data shown in Fig. \ref{fig:SUP_FMR}(a) was fitted employing a complex Lorentzian function. 

The frequency dependence of the peak-to-peak linewidth $\mu_0 \Delta$H$_{\mathrm{pp}}$ is shown in Fig. \ref{fig:SUP_FMR}(b), and the value of the Gilbert damping can be extracted using the following relation \cite{art:SUP_zakeri_FMR}:

\begin{equation}
\mu_0 \Delta \mathrm{H}_\mathrm{pp} \, = \, \frac{2}{\sqrt{3}} \frac{\alpha}{\gamma} \omega,
\end{equation}
where $\alpha$ is the Gilbert damping constant, $\gamma$ the gyromagnetic ratio, and $\omega$ the angular frequency. The value of the Gilbert damping constant was extracted by determining the slope of the frequency-dependent linewidth, and it was found to be of about $6.3 \cdot 10^{-3}$.

\begin{figure}
 \includegraphics{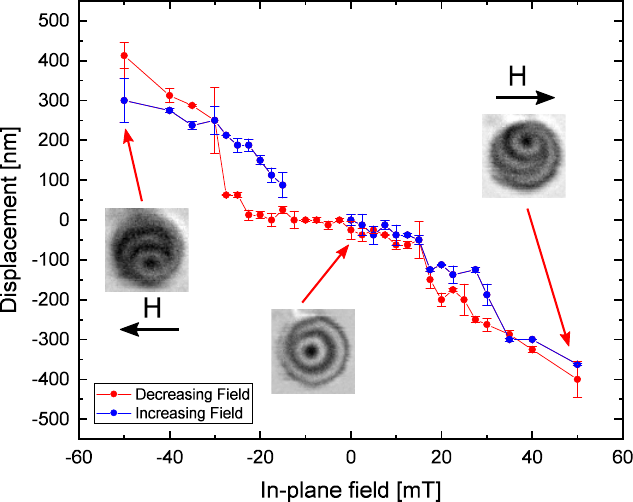}
 \caption{Position of the center of the central magnetic bubble domain at the center of a 1 $\mu$m wide thick Py disc (stabilizing a 3$\pi$ state) as a function of the external in-plane magnetic field (marked by the black arrows in the figure). A relatively smooth motion of the magnetic bubble with the external field can be observed.}
 \label{fig:bubble_displacement}
\end{figure}

To verify that the pinning sites in the Py films described here do not affect the magnetic configuration of the Py microstructured elements, we applied an in-plane magnetic field to a 1 $\mu$m diameter disc (stabilizing a 3$\pi$ state in absence of external fields). The application of an in-plane magnetic field causes the displacement of the magnetic bubble domain at the center of the structure, due to the influence of the magnetic field on the in-plane vortex state coexisting with the out-of-plane spin texture. In-plane fields of different magnitudes were applied, and a static XMCD-STXM image of the magnetic configuration of the microstructured disc was acquired at each field step. The position of the center of the magnetic bubble domain at each field step, determined from the XMCD-STXM images, is shown in Fig. \ref{fig:bubble_displacement}. A smooth displacement of the magnetic bubble with the applied field can be observed for magnetic fields below 20 mT, providing an indication that the magnetic bubble domain moves in a low pinning environment. A sharp change in the position of the magnetic bubble domain can be observed for a field of about $\pm$30 mT when approaching from the remnant state. This behavior is, similarly to what observed for the hysteresis loop of magnetic vortices \cite{art:cowburn_vortex_reversal}, to be attributed to edge repulsion effects.

\section{Dynamic properties}

In the previous section, we have shown that, by fabricating microstructured disc elements out of thick Py films, it is possible to stabilize perpendicularly magnetized magnetic configurations with different topological charges ranging from isolated magnetic skyrmions to more complex $n\pi$ states by proper selection of their diameter. Furthermore, we demonstrated that this material exhibits a low density of pinning sites and a low Gilbert damping. Thick Py films seem therefore to be ideal candidates for the investigation of the dynamical processes of topologically trivial and non-trivial configurations in PMA systems.

To demonstrate the suitability of this material for time-resolved imaging, we conducted a proof-of-principle time-resolved pump-probe measurement. In particular, we investigated the gyrotropic motion of the magnetic domain at the center of a 3 $\mu$m wide thick Py disc. As proposed in the simulations performed in Ref. \cite{art:moutafis_gyration}, the gyrotropic motion was excited by generating an oscillating out-of-plane magnetic field gradient, generated by injecting an RF current across an $\Omega$-shaped microcoil.

\begin{figure}
 \includegraphics{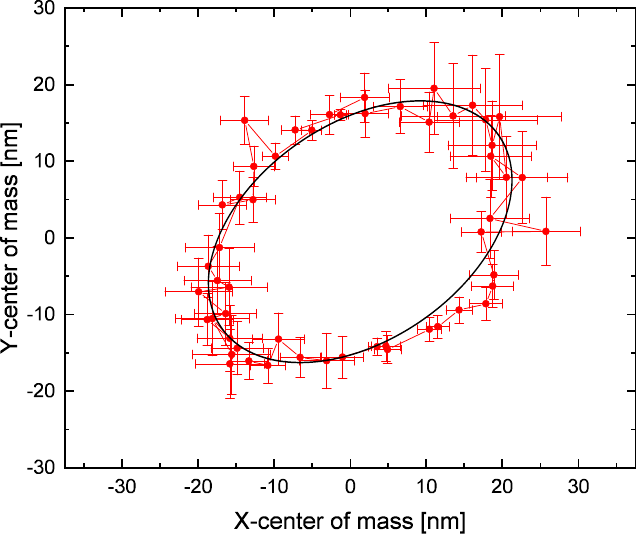}
 \caption{Position of the center of the magnetic bubble domain at the center of a 3 $\mu$m wide thick Py disc excited with an 85 MHz RF magnetic field gradient across one cycle of the RF excitation, showing an elliptical orbit. The black line acts as a guide for the eye.}
 \label{fig:dynamics}
\end{figure}

The proof-of-principle measurement was carried out by injecting RF currents with a frequency of about 85 MHz across the microcoil. The measurements were carried out in absence of externally applied static magnetic fields (remnant state). 

The center of the domain was determined for each frame of the time-resolved image by determining its magnetic center of mass. As shown in Fig. \ref{fig:dynamics}, a gyrotropic motion of the magnetic bubble domain stabilized at the center of the Py disc, with an elliptical orbit of semimajor axis of about 15 nm, can be observed in the images (a video of the time-resolved scan shown in Fig. \ref{fig:dynamics} can be found in the supplementary information). Due to the requirements of pump-probe imaging, the images shown in Fig. \ref{fig:dynamics} were acquired over 10$^8$-10$^9$ excitation cycles. This provides a demonstration of the deterministic and reproducible behavior, within the limitataions of the pump-probe imaging technique, of the time-resolved dynamics in the thick Py microstructured elements reported here, giving a final validation that thick Py films exhibiting a weak growth-induced PMA can be employed as an ideal testbed for the study of magnetodynamical processes in perpendicularly magnetized $n\pi$ spin configurations exhibiting different topological charges.

%%Conclusions

\section{Conclusions}

In conclusion, we have demonstrated that microstructured disc elements fabricated out of thick Py films grown to achieve a weak PMA stabilize, as observed in a number of previous works \cite{art:iwata_PMA_Py, art:saito_PMA_Py, art:Lo_PMA_Py, art:eames_thickPy, art:youssef_PMA_Py, art:wei_PMA_Py}, a perpendicularly magnetized configuration, composed of a circular magnetic domain at the center of the disc surrounded by ring shaped magnetic domains, the number of which is determined by the ratio between the diameter of the disc and the average width of the stripe domains \cite{art:eames_thickPy}. Depending on the diameter of the disc structures, perpendicularly magnetized states ranging from an isolated magnetic skyrmion to more complex $n\pi$ states are stabilized. Furthermore, these states are stable in absence of static out-of-plane magnetic fields.

This material exhibits both a low Gilbert damping and a low density of pinning sites. Therefore, we proposed in this work to employ this material as an ideal candidate for the investigation of dynamical processes of perpendicularly-magnetized systems, as it combines the presence of a PMA with the advantages that made Py one of the favorite materials for the study of magnetodynamical processes.

The feasibility of employing thick Py films for the study of dynamical processes in perpendicularly magnetized systems was verified by a proof-of-principle pump-probe imaging experiment, where a gyrotropic motion of the magnetic bubble stabilized at the center of a 3 $\mu$m wide Py disc was excited by an oscillating out-of-plane magnetic field gradient.

Finally, the observed weak PMA of the Py films is highly reproducible, even considering films grown in different chambers and growth conditions \cite{art:eames_thickPy, art:Lo_PMA_Py, art:youssef_PMA_Py, art:saito_PMA_Py}, providing a final reason in favor of using this material for the investigation of magnetodynamical processes in perpendicularly magnetized spin configurations.

%% If you have acknowledgments, this puts in the proper section head.

\begin{acknowledgments}
This work was performed at the PolLux (X07DA) and at the SIM (X11MA) endstations of the Swiss Light Source, Paul Scherrer Institut, Villigen, Switzerland, and at the MAXYMUS endstation of the BESSY II light source, Helmholtz Zentrum Berlin, Berlin, Germany. The research leading to these results has received funding from the European Community's Seventh Framework Programme (FP7/2007-2013) under grant agreement No. 290605 (PSI-FELLOW/COFUND), the European Union's Horizon 2020 Project MAGicSky (Grant No. 665095), and the Swiss Nanoscience Institute (Grant No. P1502). The authors thank J. Lindner for helpful discussions on the FMR results, Y. Yuan for technical support with the SQUID magnetometer, and M. Mruczkiewicz and S. Gliga for helpful discussions on the interpretation of the magnetic configuration of the Py discs.
\end{acknowledgments}

\end{document}